\documentclass[prl,twocolumn]{revtex4-1}
\usepackage{fullpage}
\usepackage{bm}        
\usepackage{amssymb}   
\usepackage{amsfonts}  
\usepackage{amsmath}   
\usepackage{mathcomp}  
\usepackage{draftcopy}
\usepackage{graphicx}
\usepackage{verbatim}
\usepackage{color}
\usepackage{ulem}
\usepackage{subfigure}

\begin{document}

\title{Overcoming the fermion sign problem in homogeneous systems}

\author{Jonathan L DuBois}
\email{jldubois@llnl.gov}
\affiliation{Lawrence Livermore National Laboratory, Livermore, CA 94550, USA}
\author{Ethan W. Brown}
\affiliation{Department of Physics, University of Illinois at Urbana-Champaign, 1110 W.\ Green St., Urbana, IL  61801-3080, USA}
\affiliation{Lawrence Livermore National Lab,  Livermore CA 94550, USA}
\author{Berni J. Alder}
\affiliation{Lawrence Livermore National Laboratory, Livermore, CA 94550, USA}

\date{\today}

\begin{abstract}
  Explicit treatment of many-body Fermi statistics in path integral Monte Carlo (PIMC) results in exponentially scaling computational cost due to the near cancellation of contributions to observables from even and odd permutations. Through direct analysis of exchange statistics we find that individual exchange probabilities in homogeneous systems are, except for finite size effects, independent of the configuration of other permutations present. For two representative systems, $^3He$ and the homogeneous electron gas, we show that this allows the entire antisymmetrized density matrix to be generated from a simple model depending on only a few parameters obtainable directly from a standard PIMC simulation.  
The result is a polynomial scaling algorithm and up to a 10 order of magnitude increase in efficiency in measuring fermionic observables for the systems considered. 
\end{abstract}

\maketitle

Path integral Monte Carlo (PIMC) methods provide essentially exact results for low temperature properties of $N$-body Bosonic systems~\cite{RevModPhys.67.279}. While the same algorithm can be applied to Fermions,  a sign problem arising from the approximately equal weights of the $N!$ oppositely signed permutations, limits the accuracy of the results. In fact, naive application of the PIMC method to fermions results in exponentially decreasing efficiency as the temperature decreases and $N$ increases \cite{ceperley103}. Consequently, enforcement of Fermi symmetry for all but the smallest finite temperature systems has so far required the introduction of an approximation that restricts path integrals to prevent sign changes, analogous to the fixed node approximation used in ground state quantum Monte Carlo (QMC)~\cite{PhysRevLett.69.331}.

\begin{figure}[b!]
\vspace{0.2 cm}
\includegraphics[width=0.35\textwidth]{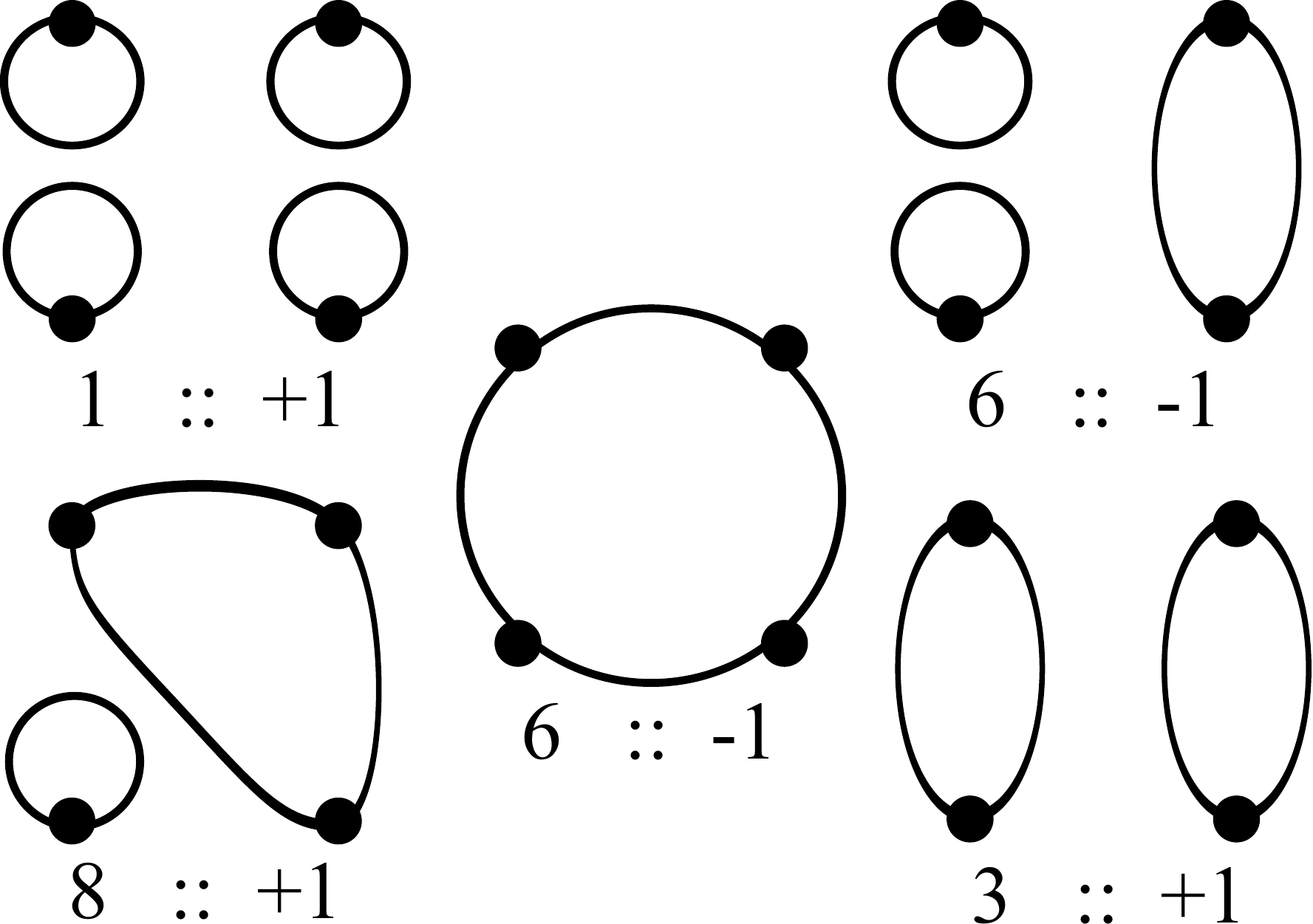}
\caption{Diagramatic representation of the equivalence classes of the symmetric group for 4 particles, $S_4$, the number of elements in each class and the sign of the contribution of members of each class to the antisymmetric partition function are shown below each diagram.}
\label{psector}
\end{figure}
In order to determine whether it is possible to overcome the sign problem directly, we have used the PIMC method to examine permutation space with great accuracy. From this data we are able to show that for homogeneous systems, the effective dimensionality of the sum over permutations can be reduced to a relatively small finite number, allowing for exact treatment of large systems to low temperatures.

The diagonal density matrix of a fermionic system at temperature $T = 1/k \beta$, denoted $\rho^{\mathcal A}({\bf R};\beta)$, can be expressed as a weighted sum over off-diagonal distinguishable density matrices, $\rho^D$, connecting the many-body coordinates, ${\bf R}$, to all permutations $\mathcal{P}({\bf R})$ in imaginary time $\beta$. This gives
\begin{equation}
\rho^{\mathcal A}({\bf R};\beta) =\frac{1}{N!} \sum_{\mathcal{P}\in S_N} (-1)^{\mathcal{P}} \rho^D({\bf R},{\mathcal{P}(\bf R)};\beta).\label{dmat}
\end{equation}
where $S_{N}$ is the symmetric group of all possible permutations of $N$ particles.

Direct treatment of this sum over permutations is made difficult in two ways: the high dimensionality of the full quantum many-body density matrix necessitates the use of a stochastic sampling method for evaluation of expectation values of observables, and the alternating sign arising from antisymmetrization over permutations of paths $(-1)^{\mathcal{P}}$ results in a large variance. Our goal in this work is to reduce the effective dimensionality of the sum over permutations.  As a starting point
along these lines we note that 
evaluation of (\ref{psector}) can be simplified by recognizing that while there are $N!$ possible permutations of $N$ particles, the symmetry group $S_N$ can be further organized into subsets of topologically equivalent diagrams $[p] \equiv \{ \mathcal{P} \in S_N | \mathcal{P} \sim p \} $~\cite{feynman-sm}. Figure \ref{psector} shows representative members from the five equivalence classes of the symmetric group for four particles, $S_4$.

Expectations evaluated within different members of the same class are identical, and one need only evaluate a single representative permutation in each class rather than the full sum. The observables for each equivalence class $[p]$ may be written as
\begin{eqnarray}
  \langle \mathcal{O} \rangle_{[p]} &=& \frac{\sum_{\mathcal{P}\in [p]}\int d{\bf R}~\hat{\mathcal{O}}\rho({\bf R},{\mathcal{P}(\bf R)};\beta)}{\sum_{\mathcal{P}\in [p]} \int d{\bf R}~\rho({\bf R},{\mathcal{P}(\bf R)};\beta)}
\end{eqnarray}
and may be measured independently during simulation.
To then reconstruct the fully antisymmetrized  observable one must sum over the equivalence classes with their respective sign,
\begin{equation}
  \langle \mathcal{O} \rangle = \frac{\sum_{[p]} \sigma_{[p]} Z_{[p]}(\beta) \langle \mathcal{O} \rangle_{[p]}}{\sum_{[p]} \sigma_{[p]} Z_{[p]}(\beta)}
  \label{observable}
\end{equation}
where we have defined $Z_{[p]}(\beta)$ as the contribution of a given equivalence class to the full symmetric partition function and $\sigma_{[p]}$ is the sign associated with the class. At zero temperature, the denominator of (\ref{observable}) becomes exactly zero. At any finite temperature, each nonequivalent permutation sector will have a different mean energy and therefore a different probability and a nonzero contribution to the antisymmetrized partition function.

The cost of summing over all unique equivalence classes of $S_N$ still scales exponentially with $N$ and attempts to make use of the structure afforded by the symmetry group directly have met with mixed success~\cite{PhysRevA.48.4075,PhysRevE.80.066702}. Ultimately, in order to obtain a polynomial scaling algorithm, it is therefore essential to determine whether all sectors need to be evaluated equally or at all. To this end, we outline a scheme to dramatically reduce the effective dimensionality of the sum over permutations.

Each permutation class is uniquely identified by the number of loops of a given length so that $[C_1,C_2,\ldots,C_N]$ represents the class with $C_1$  cycles of length one, $C_2$ cycles of length two, and so on. In a noninteracting system all sectors can be further factored into products over cycle lengths, $\ell$, allowing the contributions to the total partition function from each sector to be written as~\cite{SupplementalMaterial},
\begin{equation}
  Z_{[p]}(\beta) = \prod_{\ell=1}^{N} M_{[p]\ell} P_\ell(\beta)^{C_{[p]\ell}}
  \label{omegap}
\end{equation}
where $M_{[p]\ell} \equiv 1/(C_{[p]\ell}!\ell^{C_{[p]\ell}})$ is a combinatorial factor, and $P_\ell$ is related to the single-particle partition function and is independent of the equivalence class $[p]$. 
Given this expression for $Z_{[p]}$, observables take a particularly simple form. For example, taking the $\beta$ derivative of the partition function, one finds~\cite{SupplementalMaterial}
\begin{equation}
  \langle E(\beta) \rangle_{[p]} = \sum_{\ell=1}^N C_{[p]l} E_\ell(\beta)
  \label{energyp}
\end{equation}
where $E_\ell$ is the contribution to the toal energy from particles participating in a cycle of length $\ell$. The total energy $\langle E \rangle$ is then given by (\ref{observable}). A similar expression can be constructed for the pair correlation function and other observables \cite{SupplementalMaterial}.  

The significance of (\ref{omegap}) and (\ref{energyp}) within the context of the sign problem in PIMC is that expectation values can be obtained by evaluating the $N$ positive definite expectations $P_\ell(\beta)$ rather than the $N!$ terms in (\ref{dmat}).
The net result is an algorithm with an effective computational cost scaling as well as $\mathcal{O}(N^2)$ in the number of particles since the probability density associated with each of the equivalence classes can be reconstructed from the probability densities of an $\mathcal{O}(N)$ subset of $S_N$ (e.g. $[N,0,0,\ldots], [0,N/2,0,0,\ldots], [0,0,N/3,0,\ldots], etc.$) and the computational cost of the PIMC algorithm itself can be $\mathcal{O}(N)$.

Going further, it can be shown that for a noninteracting (ideal) gas the contribution to the partition function of neighboring sectors is proportional to their length~\cite{SupplementalMaterial} allowing one to relate $P_\ell$ to an exponentially decreasing function in cycle length so that $P_\ell = p_2^{-\ell}$. This single expectation, $p_2$, can be seen as the mean probability of a permutation between two particles.
Taking advantage of these observations allow for a reduction of the task of finding the relative probability of different sectors to that of determining a single temperature dependent value, leading to an $O(N)$ algorithm. 

In an interacting system one might assume that the relative probability of different cycle lengths $P_\ell$ will depend on the permutation sector $[p]$ and construct a sector dependent $P_{[p]\ell}$ as an expansion around an averaged $\bar{P}_\ell$ such that 
\begin{eqnarray}
  P_{[p]\ell}/\bar{P}_\ell = &1 + \sum_{m=1}^{N} C_{[p]m}(\frac{\bar{P}_{\ell m}}{\bar{P}_\ell} -1)+ \nonumber \\
&\sum_{m,n=1}^{N} C_{[p]m}C_{[p]n}(\frac{\bar{P}_{\ell mn}}{\bar{P}_\ell} -1)+ \ldots
\label{ClusterExpansion}
\end{eqnarray}
where $\bar{P}_{\ell m\ldots}$ is related to the probability of finding a cycle length $\ell$ given the existence of cycles of lengths $m\ldots$.  Krauth and Holzmann have shown that for weakly interacting bosons effective interactions between different cycle lengths, i.e.  $\bar{P}_{\ell m\ldots}$, are negligible \cite{PhysRevLett.83.2687,krauth-smac} suggesting that determination of $\bar{P}_\ell$ alone is sufficient for a wide class of systems~\footnote{The idea that the qualitative features of the loop structure may be obtained from the pair exchange probability alone was already proposed by Feynman \cite{PhysRev.91.1291} to describe the $\lambda$ transition in liquid $^4$He. In order to evaluate the weight of each permutation sector analytically, Feynman assumed that the number of ways of forming a closed loop consisting of $k$ particles is approximately independent of the arrangement of other loops and could be accounted for by a single effective parameter. This results in the relative probability of different permutation sectors being determined by a simple Poisson distribution}.  In this work we have found through direct PIMC simulation that this same simple structure is obeyed in two important strongly interacting fermi systems. Our key finding is that, aside from small finite size effects, the contribution of a loop of length $\ell$ to the probability of a permutation sector is the same for all permutation sectors. This remarkable result leads to a dramatic reduction in the computational cost required to evaluate any signed observable. 

In what follows, we present results of this approach applied to two prototypical strongly interacting fermonic liquids -- the homogeneous electron gas (HEG) and liquid $^3$He.   The character of exchange interactions in these two systems represent two qualitative extremes.  In the HEG, a weak correlation hole results in a high probability of exchange between nearest neighbors.  In contrast, $^3$He has a strong correlation hole resulting in a significantly lower nearest neighbor exchange probability.  As a consequence of its higher compressibility and higher exchange probability, we find that the permutation structure of the HEG more closely resembles the noninteracting gas and the average value of the sign in the HEG, $\langle \sigma \rangle$, decays to zero more rapidly with decreasing temperature and increasing number of particles than in $^3$He. The strong correlation hole in $^3$He has the effect of modifying the combinatorial factor, $M_{[p]\ell}$, in (\ref{omegap}) for a finite simulation box requiring the addition of a model to account for the absence of overlapping exchange loops. 
\begin{table*}[t]
  \begin{tabular}{|c|c|c|c|c|c|}
    \cline{1-6}
    $r_{s}$ & $T/T_{F}$ & $PIMC$ & $RPIMC$ & $P_{l}$ & $p_{2}$ \\
    \cline{1-6}
    $1.0$ & $0.125$ & $1(10)$ & $2.35(1)$ & $2.3(1)$ & $2.33(6)$ \\
    $1.0$ & $1.0$ & $3(7)$ & $8.69(3)$ & $8.7815(7)$ & $8.7801(7)$ \\
    $10.0$ & $0.125$ & $-0.0(1)$ & $-0.1038(2)$ & $-0.1030(1)$ & $-0.1033(1)$ \\
    $10.0$ & $1.0$ & $-0.040(2)$ & $-0.0403(5)$ & $-0.0402(1)$ & $-0.04025(5)$ \\
    \cline{1-6}
  \end{tabular}
  \caption{Total energies per particle for $33$ spin-polarized electrons at $r_s=1$, $10$ and $T/T_F=0.125$, $1.0$. From left to right, we plot energy estimates for standard signful PIMC, restricted PIMC from \cite{PhysRevLett.110.146405}, reconstructed PIMC using $P_\ell$ as in (\ref{omegap}), and reconstructed PIMC using $p_2$. }
  \label{hegenergies}
\end{table*}

We have examined both a low density ($r_s = 10$) and high density ($r_s = 1$) state of the HEG both at the Fermi temperature, $T_F$, and at $1/8$ $T_F$. Table (\ref{hegenergies}) summarizes energies of the HEG for these two densities and temperatures compared with previous exact and fixed-node results \cite{PhysRevLett.110.146405}. We find that our reconstructed energies, both by fitting $p_2$ and $P_\ell$ directly, match well with previous fixed-node results.  This agreement provides a direct demonstration that the permutation structure of the HEG is well described by free fermions with an effective mass. 
Results obtained by fitting $P_\ell$ are within error bars of those obtained with the more constrained $p_2$ fits.  Previous exact simulations were not possible below the Fermi temperature, giving estimates for the energy with a variance larger than the value itself, while both reconstructions work well at $1/8$ $T_F$.  Our new method thus extends the regime where unbiased exact simulations are possible to much lower  temperatures.  The increase in efficiency is most notable for $rs=1.0$ and $T/T_F = 1.0$ where resummation of the same PIMC data using the $p_2$ model results in a standard error 5 orders of magnitude smaller than the direct antisymmetrized sum.  Given that the statistical error scales with the square root of the number of independent samples, an additional $10^{10}$ times as many samples would be required to obtain the same result by direct summation.
 For some of the points examined, we see up to $\sim 1\%$ discrepancies with the fixed-node result. It is tempting to assume that the current results are more accurate since they avoid the fixed node approximation. However, in order to confirm this one would need to include higher order terms in the cluster expansion (\ref{ClusterExpansion}) and demonstrate convergence.    

\begin{figure}
\includegraphics[width=0.45\textwidth]{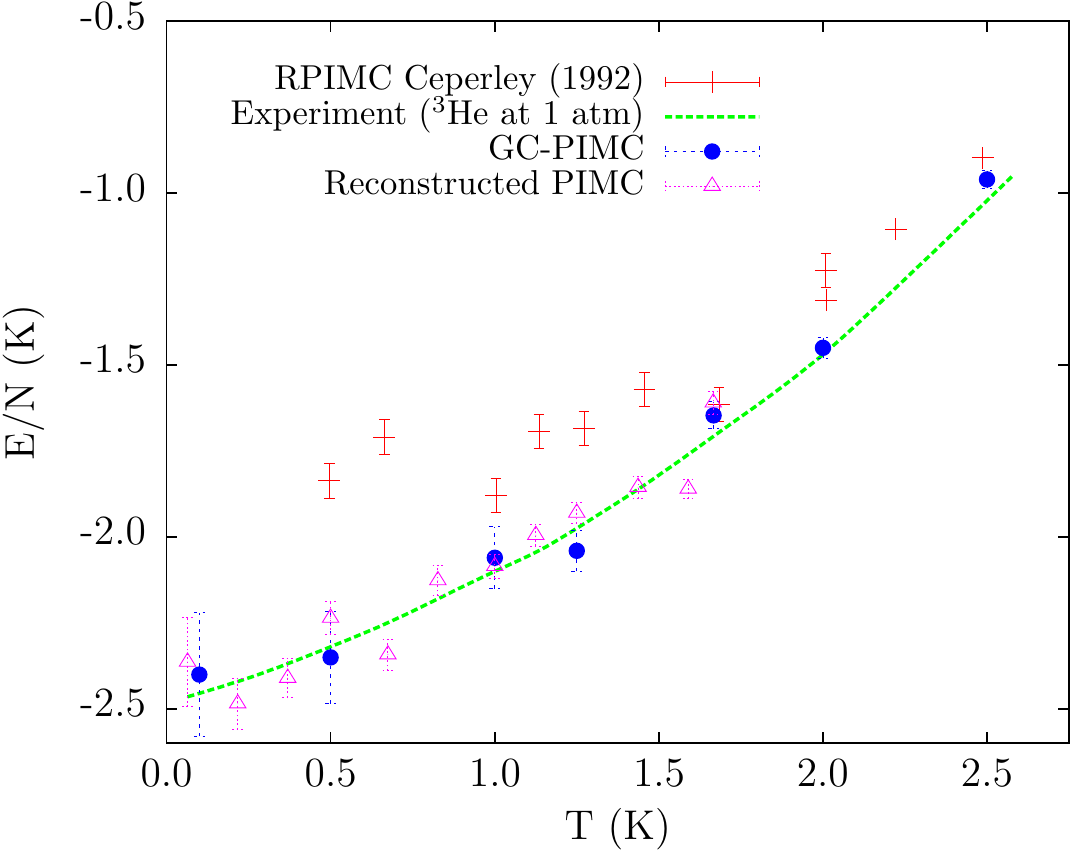}
\caption{Results of antisymmetrized grand canonical PIMC applied to liquid $^3$He. Our direct results (solid circles) and the reconstructed energies described in the text (open triangles) agree well with experimental data (dashed line) down to temperatures well below the $^3$He Fermi temperature.  Results obtained with restricted PIMC in~\cite{PhysRevLett.69.331} (+ signs) are shown for comparison. }
\label{result}
\end{figure}

Results of our approach applied to liquid $^3$He are shown in Figure~\ref{result}. An approximate (truncated) direct summation over permutations performed in a previous work~\cite{pacifichem10} (solid circles) agrees well with experimental data~\cite{PhysRevB.27.2747} (dashed line) down to temperatures well below the $^3$He Fermi temperature, $T_f=1.7 K$.  Reconstructed energies were obtained with an $\approx 10$ order of magnitude reduction in computational cost compared to a naive direct summation over partitions and agree to within statistical errorbars with the experimental values as opposed to the approximation introduced by the restricted path method~\cite{PhysRevLett.69.331}.

The strong correlation hole in $^3$He results in a deviation of the combinatorial factor $M_{[p]\ell}$ in (\ref{omegap}) in finite systems from the free gas (overlapping permutation loops) model. In order to account for this and extract a thermodynamic limit value for $p_2$ from our finite-$N$ PIMC data, we numerically solve an analogous but discrete problem namely, a nearest neighbor Ising model with $N$ sites and a nearest neighbor connectivity close to that of the liquid for each distinct permutation sector. \footnote{We note that the choice for the Ising lattice is somewhat ad hoc in the sense that if a square lattice is assumed, where each He atom has 6 nearest neighbors whereas an e.g. hcp lattice would have 12 neighbors.  However, we have found that while the choice of lattice does change the mean extracted value of $p_2$ somewhat,  the reconstructed values of $Z_\mathcal{P}$ do not depend significantly on the choice of lattice, being insensitive to the detailed form of the lattice for the temperatures we have considered.}

\begin{figure}
\includegraphics[width=0.45\textwidth]{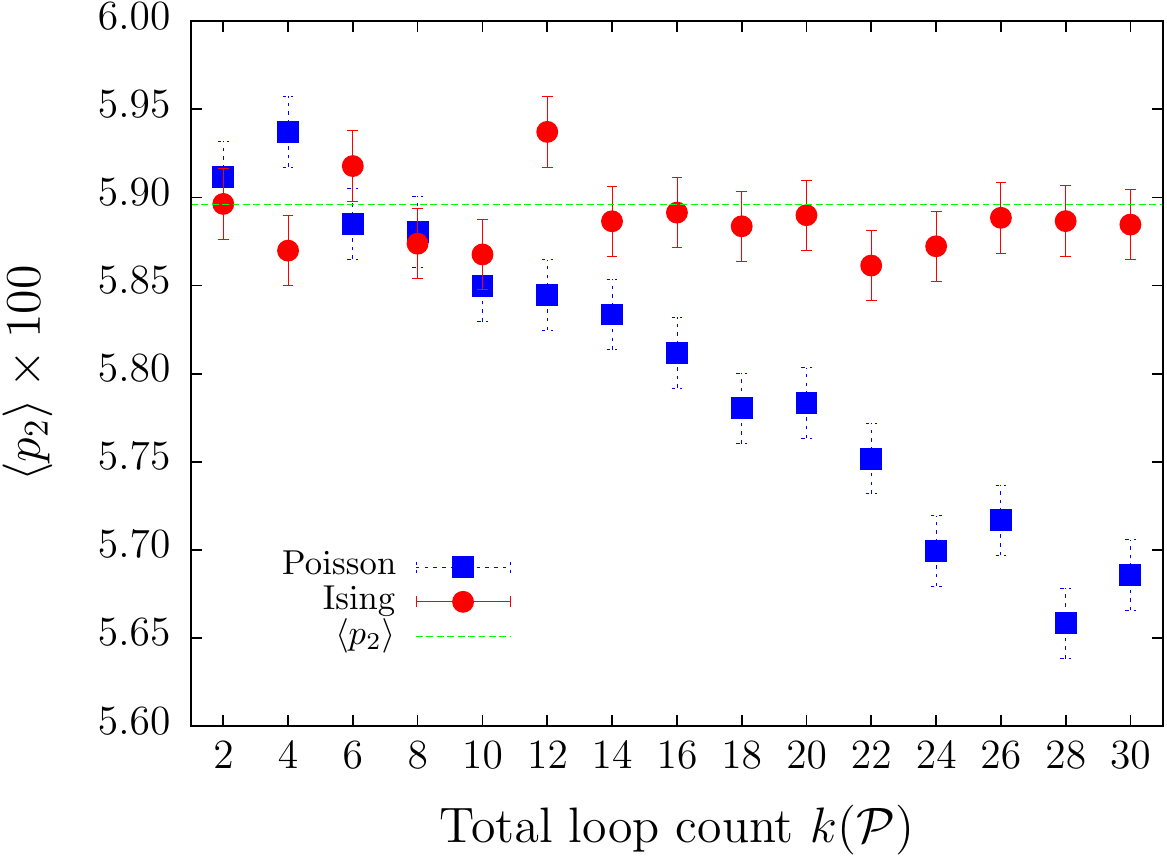}
\caption{Representative values of the extracted pair exchange probability $p_2$ across a representative subset of equivalence classes containing increasingly long loops using Poisson statistics (squares) and the numerically inverted Ising model (circles).  The dashed line shows the mean value $\langle p_2 \rangle$ over all sectors obtained via the Ising method.}
\label{p2plot}
\end{figure}

Figure \ref{p2plot} shows the values of $p_2$ for a range of permutation sectors for $N=66$ $^3$He at $T=1.2$ Kelvin. Solid squares in the figure show results for $p_2$ obtained based on an uncorrected Poisson model. The statistically significant drift in the mean value of $p_2$ with increasingly long permutation cycles resulting from finite size effects is evident. In contrast, results for $p_2$ obtained from the inverse Ising model are found to agree with the mean value obtained over all sectors within statistical error bars ($\approx 1\%$). The mean value of $p_2$ obtained in this way was used to weight expectation values over the full antisymmetrized density matrix.

In conclusion, we have shown that it is possible to directly address the sign problem for homogeneous systems by taking advantage of their relatively simple permutation space structure. While here we report only energies, we note that other diagonal observables may be accessed in the same way \cite{SupplementalMaterial}. Current efforts are focused on reconstructing the permutation space of heterogeneous systems in the hope that we can also exactly calculate properties of general many-body Fermion systems much below the Fermi temperature.

\begin{acknowledgements}
This work was performed under the auspices of the U.S. Department of Energy by Lawrence Livermore National Laboratory under Contract No. DE-AC52-07NA27344 and supported by LDRD Grant No. 10-ERD-058 and the Lawrence Scholar program.
\end{acknowledgements}

\bibliography{fpimc-14.bib}

\onecolumngrid
\setcounter{equation}{0}
\setcounter{figure}{0}
\setcounter{table}{0}
\setcounter{page}{1}
\makeatletter
\renewcommand{\theequation}{S\arabic{equation}}
\renewcommand{\thefigure}{S\arabic{figure}}
\renewcommand{\bibnumfmt}[1]{[S#1]}
\renewcommand{\citenumfont}[1]{S#1}

\section{Supplemental Material: Overcoming the fermion sign problem in homogeneous systems}

\centerline{\bf Jonathan L DuBois$^{\dagger}$, Ethan W. Brown$^{\dagger \star}$ and  Berni J. Alder$^{\dagger}$}
\centerline{$^\dagger$ Lawrence Livermore National Laboratory, Livermore, CA 94550, USA}
\centerline{$^\star$ Department of Physics, University of Illinois at Urbana-Champaign,}
\centerline{ 1110 W.\ Green St., Urbana, IL  61801-3080, USA}

\hspace{1cm}

In the proceeding supplemental material we provided a derivation of the permutation space structure of the free Fermi gas. We explicitly show how the entire $N!$ space may be written in terms of $N$ variables or, in some limiting cases, a single parameter and discuss the impact on the fixed cost statistical error in the mean value of the sign. Finally we show how this affects various diagonal observables, namely the total energy and pair correlation function. 

\section{Fermion sign problem and error propagation for permutation models}

We begin by writing the many-body, interacting quantum-statistical partition function of $N$ identical particles as
\begin{eqnarray}
  \label{eq:mbz}
  \mathcal{Z}^{B/F}(\beta) &=& Tr(\rho^{B/F}) \\
                           &=& \frac{1}{N!}\sum_{\mathcal{P}\in\mathcal{S}_{N}}\sigma^{B/F}_{\mathcal{P}} \int d{\bf R} \rho^D({\bf R},{\mathcal{P}(\bf R)};\beta)
\end{eqnarray}
where the sum is over all possible permutations $\mathcal{P}$ belonging to the symmetric group $\mathcal{S}_{N}$ for $N$ particles. The labels $B,F,D$ signify that these are boson, fermion, and distinguishable (Boltzmannon) quantities respectively. The weight $\sigma_{\mathcal{P}} \in \{-1,+1\}$ for each respective permutation depends on its constituent particle statistics / symmetry. For bosons, all permutations contribute positive weight, making $\rho^{B}$ a real probability distribution. For fermions, however, only permutations with an even number of exchanges contribute positive weight, while those with an odd number of exchanges contribute negative weight. Direct Monte Carlo sampling, however, requires a kernel that is a positive definite and bounded probability distribution.  To accommodate this, Path integral Monte Carlo (PIMC) sampling of the antisymmetric density matrix typically samples the bosonic density matrix (i.e explicit sampling of permutations but no sign change for odd and even permutations) and the sign arising from fermi symmetry is kept as a weight applied to each measurement. This results in fermionic observables take the form
\begin{eqnarray}
  \langle\hat{\mathcal{O}}\rangle_{F} & = & Tr(\hat{\mathcal{O}}\rho^{F})/Tr(\rho^{F})\\
                                      & = & \frac{Tr(\hat{\mathcal{O}}\rho^{F})/Tr(\rho^{B})}{Tr(\rho^{F})/Tr(\rho^{B})}\\
                                      & = & \langle\hat{\mathcal{O}}^{F}\rangle_{B}/\langle\sigma^{F}\rangle_{B}
  \label{eq:SignedObservableFermion}
\end{eqnarray}
where we now need to collect statistics on both the signed observable $\hat{\mathcal{O}}^{F}$ and average sign weight $\sigma^{F}$ while sampling from the bosonic distribution. The denominator of (\ref{eq:SignedObservableFermion}) can be rewritten as
\begin{eqnarray}
  \label{eq:sgn}
  \langle\sigma^{F}\rangle_{B} = \sum_{\mathcal{P}} \sigma^{F}_{\mathcal{P}} \mathcal{Z}^{D}_{\mathcal{P}}(\beta) = \frac{\mathcal{Z}^{F}(\beta)}{\mathcal{Z}^{B}(\beta)} = \exp{[-\beta(\mathcal{F}^{F}(\beta) - \mathcal{F}^{B}(\beta))]}
\end{eqnarray}
where $\mathcal{F}$ is the many-body free energy and $\mathcal{Z}^{D}_{\mathcal{P}}(\beta) = \int d{\bf R} \rho^D({\bf R},{\mathcal{P}({\bf R})};\beta)$ is the contribution to the full partition function resulting from integration over the off diagonal distinguishable particle density matrix for an explicit permutation $\mathcal{P}$. The fermion free energy is always greater than or equal to the boson free energy with the difference between the two growing as temperature is reduced. This free energy difference is most striking for non-interacting particles where from (\ref{eq:sgn}) we find the average value of the sign becomes
\begin{equation}
  \langle \sigma^{F} \rangle_{B,0} = \exp{[-\frac{N^2}{V}(2\pi\lambda\beta)^{D/2}]}.
  \label{eq:SignFree}
\end{equation}
For the noninteracting fermi gas it is thus clear that the average value of the sign decreases exponentially with decreasing temperature and increasing number of particles.   In order to understand how the decreasing magnitude of the average sign impacts the computational cost associated with obtaining a fermionic expectation value we must examine how it impacts the statistical error within Monte Carlo. 

The standard error of the mean value of the sign is 
\begin{equation}
  \epsilon(\sigma^{F}) \equiv \sqrt{\frac{var(\sigma^{F})}{M}}
\end{equation}
where $var$ represents statistical variance and $M$ is the number of independent samples. Using (\ref{eq:sgn}) we can write $var(\sigma^{F})$ in terms of the variances of each permutation sector weight. If we assume the permutation sector weights are uncorrelated, we arrive at
\begin{equation}
  \epsilon(\sigma^{F}) = (\sum_{\mathcal{P}} \frac{var(\mathcal{Z}_{\mathcal{P}}^{D})}{M})^{1/2} = (\sum_{\mathcal{P}} \epsilon(\mathcal{Z}_{\mathcal{P}}^{D})^{2})^{1/2} \propto |S(N)|\epsilon(\mathcal{Z}_{\mathcal{P}}^{D}).
\end{equation}
Given that all observables measured in simulation must be normalized by $\langle\sigma\rangle$, there is a corresponding exponential increase in the variance of signed observables $\hat{\mathcal{O}}_{F}$, causing their precise calculation to be computationally intractable. This is the signature of the \textit{fermion sign problem}.

Alternatively, suppose each $\mathcal{Z}_{\mathcal{P}}^{D}$ is not independent, but may be generated by a function of $N$-parameters
\begin{equation}
  \mathcal{Z}_{\mathcal{P}}^{D} = f(b_{1},\dots,b_{N};\mathcal{P})
\end{equation}
This leads to
\begin{equation}
  \epsilon(\mathcal{Z}_{\mathcal{P}}^{D})_{f} = (\sum_{i=1}^{N} \frac{1}{M_{b_{i}}} (\frac{\partial f}{\partial b_{i}})^{2} var(b_{i}))^{1/2}
\end{equation}
where $M_{b_{i}}$ is the number of statistically independent samples of each variable $b_{i}$. Comparing these two definitions of the error of each permutation sector weight, we find
\begin{equation}
  \frac{\epsilon(\mathcal{Z}_{\mathcal{P}}^{D})_{f}}{\epsilon(\mathcal{Z}_{\mathcal{P}}^{D})} = (\sum_{i=1}^{N} \frac{M}{M_{b_{i}}} (\frac{\partial f}{\partial b_{i}})^{2} \frac{var(b_{i})}{var(\mathcal{Z}_{\mathcal{P}}^{D})})^{1/2}.
\end{equation}
Since each permutation sector weight is written in terms of the parameters $b_{i}$, each sample of $\mathcal{Z}_{\mathcal{P}}^{D}$ is also a sample of $b_{i}$. Furthermore, since we assume $b_{i}$ is independent of permutation sector, if all $N!$ sectors are sampled $M$ times, the factor $M_{b_{i}}$ is effectively $N!M$. This leads to a dramatic reduction in the error associated with each permutation sector weight,
\begin{equation}
  \frac{\epsilon(\mathcal{Z}_{\mathcal{P}}^{D})_{f}}{\epsilon(\mathcal{Z}_{\mathcal{P}}^{D})} \sim \frac{1}{\sqrt{(N-1)!}},
\end{equation}
and a corresponding reduction in the error of the average Fermi sign. In the following we motivate several such models $f(\{b_{i}\};\mathcal{P})$. Empirical demonstration of their effectiveness is given in the main text.

\section{Separating permutation space}

\subsection{Permutation sectors}

\begin{figure}[b!]
\includegraphics[width=0.35\textwidth]{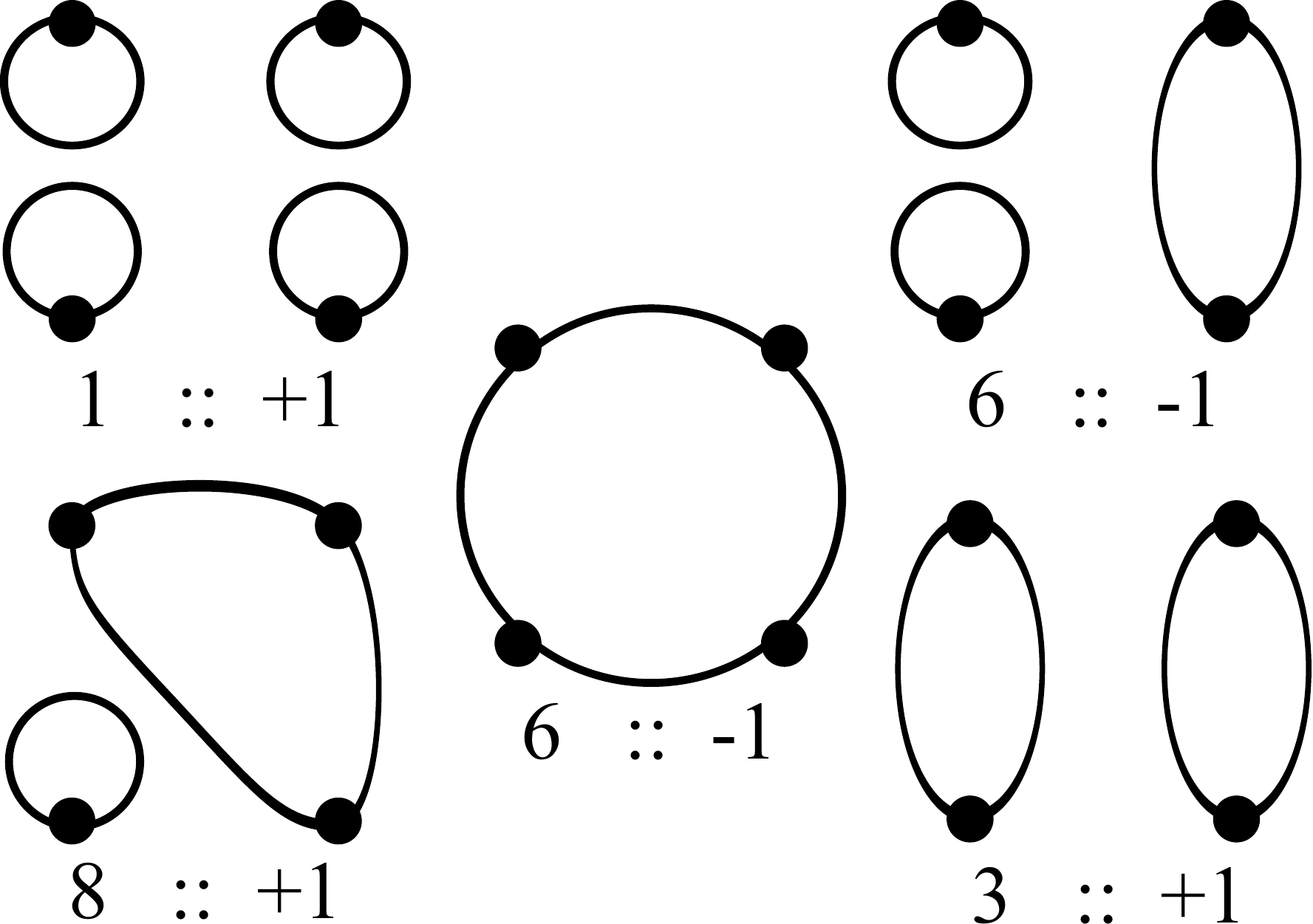}
\caption{Diagramatic representation of the equivalence classes of the symmetric group for 4 particles, $S_4$, the number of elements in each class and the sign of the contribution of members of each class to the partition function are shown below each diagram.}
\label{psector}
\end{figure}

To arrive at an effective model of permutation space, the first step is to notice that permutation space may be divided into permutation classes $[p] \equiv \{ \mathcal{P} \in S_N | \mathcal{P} \sim p \}$. These sectors may be defined explicitly by the number of different length cycles they contain. (For an example with $N=4$ see Fig. \ref{psector}.) The number of sectors in $[p]$ consisting of $C_{[p]1} \dots C_{[p]N}$ cycles of length $l_{1} \dots l_{N}$ is
\begin{equation}
  M_{[p]} = \frac{N!}{\prod_{l=1}^{N} C_{[p]l}! l^{C_{[p]l}}}.
\end{equation}
This allows us to write the full $N$-particle partition function
\begin{equation}
  \mathcal{Z}^{(N)}(\beta) = \frac{1}{N!} \sum_{[p]} \sigma_{[p]} M_{[p]} \mathcal{Z}^{(N)}_{[p]}(\beta)
\end{equation}
where we have defined the $N$-particle partition function of a single permutation sector as
\begin{equation}
  \mathcal{Z}^{(N)}_{[p]}(\beta) \equiv \int d{\bf R} (\prod_{i=1}^{N} \int_{r_{i}(0)}^{r_{[p](i)}(\hbar \beta)} d r_{i}) \exp^{-\mathcal{A}/\hbar}.
\end{equation}
Here the path of particle $i$ goes from $r_{i}(0)$ to $r_{[p](i)}(\hbar\beta)$, and $\mathcal{A}$ represents the many-body action. This provides a dramatic reduction in the number of independent terms in the antisymmetric sum. For example, for $N=33$ particles the sum is reduced from $\sim 10^{38}$ to $\sim 10^{5}$ terms. Nevertheless, efficiency still scales exponentially in the number of particles. To make further progress, we turn to the free Fermi gas.

\subsection{Free Fermi gas}

For free particles, the integral over the exponentiated action can be performed analytically, leading to
\begin{eqnarray}
  \mathcal{Z}^{(N)}_{[p]}(\beta) &=& \int d{\bf R} \prod_{i=1}^{N} \frac{1}{\sqrt{4\pi\lambda\beta}^{D}} \exp{[-\frac{(r_{[p](i)}-r_{i})^{2}}{4\lambda\beta}]}.
\end{eqnarray}
where we define $\lambda \equiv \frac{\hbar^{2}}{2m}$. Each permutation sector $[p]$ decomposes into mutually disconnected groups, each with a winding number $l$, subject to the constraint that $N = \sum_{l}l C_{[p]l}$. For each group, we can then take each integral separately, leading to
\begin{equation}
  \label{freegaspf}
  \mathcal{Z}_{0}^{(N)}(\beta) = \frac{1}{N!} \sum_{[p]} \sigma_{[p]} M_{[p]} \prod_{l=1}^{N} \mathcal{Z}_{0}^{(1)}(l\beta)^{C_{[p]l}}.
\end{equation}
This final form shows that all $N!$ permutation sectors may be written in terms of $N$ parameters, specifically the free one particle partition function at inverse temperature $l\beta$.

Generally we can write the free one particle partition function in a box of volume $\Omega \equiv L^{D}$ as
\begin{equation}
  \mathcal{Z}_{0}^{(1)}(\beta) = \sum_{n} \exp{[-\beta \lambda (\frac{2\pi n}{L})^2]} = \frac{\Omega}{\sqrt{4\pi\lambda\beta}^{D}} \sum_{n} \exp{[-\frac{(nL)^{2}}{4\lambda\beta}]}
\end{equation}
where $n \equiv (n_1,\dots,n_D)$ with $n_{i} = -\infty,\dots,\infty$. These two forms are related by a Laplace transformation. In the limit $L \gg \lambda$, the sum over states may be approximated as an integral leading to
\begin{equation}
  \mathcal{Z}_{0}(l\beta) = \frac{\Omega}{\sqrt{4\pi\lambda l\beta}^{D}} = (\frac{\gamma}{l})^{D/2} \mathcal{Z}_{0}(\gamma\beta).
\end{equation}
Thus in the thermodynamic limit, the entirety of permutation space may be formed from a single parameter, $\gamma$.

\subsection{Interacting Fermi gas}

We now turn our attention to what happens when we include interactions. In the mid 20th century Feynman and Kikuchi presented a model for the lambda transition in liquid $^{4}$He \cite{feynman-sm}. Through the use of physical intuition granted by the path integral representation, Feynman was able to show the effect of the potential during a permutation is only to change the effective mass, $\lambda \rightarrow \lambda'$. This allows one to write down the many-body partition function as
\begin{equation}
  \label{FeynmanPF}
  \mathcal{Z}^{(N)}(\beta) = \frac{1}{N!} \sum_{[p]} \sigma_{[p]} M_{[p]} \int d{\bf R} \rho(r_{1},\dots,r_{N};\beta) \prod_{i=1}^{N} \frac{1}{\sqrt{4\pi\lambda'\beta}^{D}} \exp{[-\frac{(r_{[p](i)}-r_{i})^{2}}{4\lambda'\beta}]}
\end{equation}
where the first term $\rho$ is the potential's contribution to the initial configuration. Thus, in line with a free gas, we may assume that exchanges are essentially independent, allowing us to write the many-body partition function in terms of exchange frequencies $P_l$. With this in mind, we may write
\begin{equation}
  \label{PlModel}
  \mathcal{Z}^{(N)}(\beta) = \frac{1}{N!} \sum_{[p]} \sigma_{[p]} M_{[p]} \prod_{l=1}^{N} P_{l}(\beta)^{C_{[p]l}}.
\end{equation}
This bares a remarkable resemblance to the free particle partition function of (\ref{freegaspf}), except here $Z_{0}^{(1)}(l\beta)$ has been replaced by an exchange frequency $P_{l}(\beta)$ which embeds the effect of interactions. Thus, we have again reduced the required simulation space from $N!$ to $N$.

Feynman and Kikuchi take this a step further, proposing an intuitive model for $P_{l}(\beta)$. Assuming there is a well-defined mean distance of permuting $d$, one can use it to replace $(r_{[p](i)}-r_{i})$ in (\ref{FeynmanPF}). This allows us to move the exponential term outside of the integral, suggesting the form
\begin{equation}
  P_{l}(\beta) = \exp{[-\frac{d^{2}l}{4\lambda'\beta}]}
\end{equation}
which again reduces the problem to a single unknown parameter. In the main article, we rewrite this as $P_{l}(\beta) = p_{2}^{-l}(\beta)$ and attempt to uncover the value of $p_{2}$, i.e. the probability of a pair-wise exchange.

\section{Simulation details}

When presenting the above analysis, Feynman mentions, ``The above results apply only to Bose particles. Fermi particles behave differently'' \cite{feynman-sm}. We note however that in PIMC simulation of fermions, one actually simulates bosons and only keeps track of the current overall sign as a weight. At the end of the simulation, we divide all observables by the expectation value of this weight, i.e. the average value of the sign $\langle \sigma \rangle$. As explained above, the variance in $\langle \sigma \rangle$ is the source of the sign problem, so any new approach to the sign problem must address it.

\begin{figure}
\centering
\begin{subfigure}
  \centering
  \includegraphics[width=.45\linewidth]{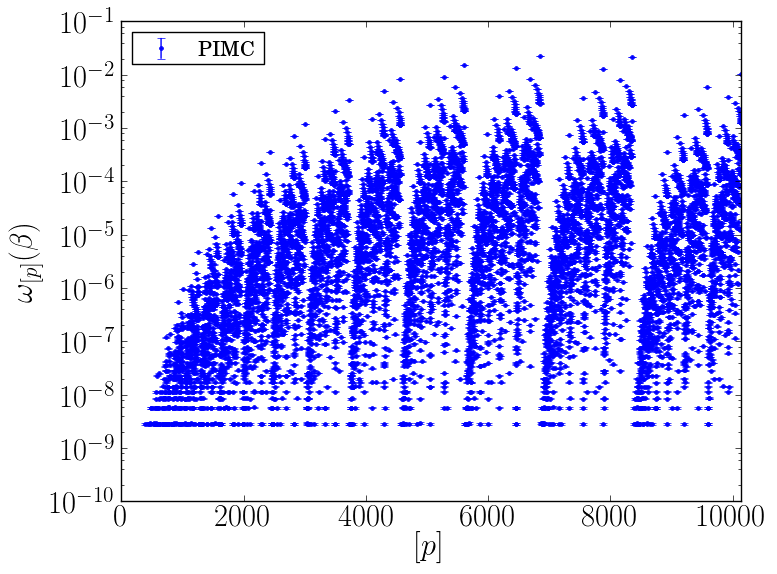}
  \label{permprobsA}
\end{subfigure}%
\begin{subfigure}
  \centering
  \includegraphics[width=.45\linewidth]{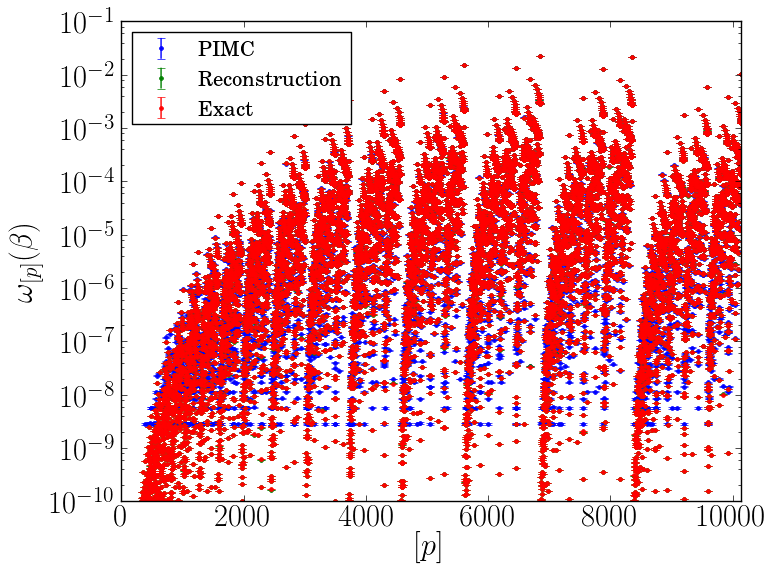}
  \label{permprobsB}
\end{subfigure}
\caption{Permutation sector probabilities $\omega_{[p]}(\beta)$ for a free Fermi gas. The blue points shown in both plots are from PIMC simulation. At the lowest probabilities the blue points form straight lines representing visiting those sectors $[p]$ only $1, 2, 3, \dots$ times. The green points represent reconstructed values using the model (\ref{PlModel}). However, these points are perfectly masked by the exact analytical values, shown in red. One can see the reconstruction extends the permutation structure to orders of magnitude lower probabilities.}
\label{permprobs}
\end{figure}

The first step in our approach to improving this variance is to measure and examine the probability of each permutation sector $\omega_{[p]}(\beta)$ as defined in the main text. What we find is not so surprising: If the average value of the sign is on the order of or less than the lowest probability sectors, the variance of the sign is larger than its value. Our goal then is bring this line down below the order of the sign. Naively this is accomplished by simply running the simulation for longer and collecting more statistics, however as mentioned this process scales exponentially. Instead we propose a new technique wherein the sectors with already good statistics are used to precisely fit the aforementioned models, moving from an $\mathcal{O}(N!)$ space to an $\mathcal{O}(N)$ or even $\mathcal{O}(1)$ one. This fitted model is then used to reconstruct the rest of the lower probability permutation sectors, moving the line well below that of the average value of the sign, see Fig. \ref{permprobs}.

For $^3$He, we have utilized the grand-canonical worm algorithm which has been shown to efficiently sample permutation space. Additionally, we have used standard importance-sampling techniques to deemphasize the contribution of long permutation cycles in the sum over permutations. In this approach, sampling of long permutation cycles is penalized by reducing the probability of attempting moves that will extend the length of a permutation while the acceptance probability of such moves is increased to maintain detailed balance. The net result is that permutations with a few short permutation cycles are sampled often and high orders rarely. For $N=66$ unpolarized $^3$He sampling only relevant sectors reduces the number of terms in the sum over permutations from $33!^2$ to $\approx 6000$~\cite{pacifichem10}.

\section{Observables}

Now that we have presented models for the many-body partition function, it is worthwhile to see how this affects various observables we might want to measure. Here we focus on two diagonal observables: the total energy and the pair-correlation function.

\subsection{Total energy}

Recall the energy may be defined as,
\begin{equation}
  E^{(N)}(\beta) = \frac{\partial}{\partial\beta} \log{\mathcal{Z}^{(N)}(\beta)} = \frac{1}{\mathcal{Z}^{(N)}(\beta)} \frac{\partial}{\partial\beta} \mathcal{Z}^{(N)}(\beta).
\end{equation}
For the model presented in (\ref{PlModel}) then,

\begin{eqnarray}
  E^{(N)}(\beta) &=& \frac{1}{\mathcal{Z}^{(N)}(\beta)} \frac{\partial}{\partial\beta} [\frac{1}{N!} \sum_{[p]} \sigma_{[p]} M_{[p]} \prod_{l=1}^{N} P_{l}(\beta)^{C_{[p]l}}] \\
                 &=& \frac{1}{\mathcal{Z}^{(N)}(\beta)} \frac{1}{N!} \sum_{[p]} \sigma_{[p]} M_{[p]} \prod_{l=1}^{N} P_{l}(\beta)^{C_{[p]l}} \sum_{\gamma=1}^{N} C_{[p]\gamma} \frac{\frac{\partial}{\partial\beta}P_{\gamma}(\beta)}{P_{\gamma}(\beta)} \\
                 &=& \sum_{[p]} \sigma_{[p]} \omega_{[p]}(\beta) \sum_{l=1}^{N} C_{[p]l} E_{l}(\beta)
\end{eqnarray}
where we have defined $\omega_{[p]}(\beta) \equiv \frac{M_{[p]}}{N! \mathcal{Z}^{(N)}(\beta)}$ and $E_{l}(\beta) \equiv \frac{\frac{\partial}{\partial\beta}P_{l}(\beta)}{P_{l}(\beta)}$ which are the permutation sector probability and cycle length energy respectively. This demonstrates that if the permutation sector probabilities can be described by $N$ parameters, so may the energy. Similarly to reconstructing the sign, we can take those sectors for which we have good estimates of the sector energy and use it to fit the $N$ $E_{l}$'s and from them rebuild all $N!$ $E_{[p]}$'s.

\subsection{Pair correlation function}

Finally recall the pair correlation function is defined as,
\begin{equation}
  g(r) = \frac{2 \Omega}{N^{2}} \langle \sum_{i<j} \delta(r^{(i)}-r^{(j)}-r) \rangle.
\end{equation}
According to (\ref{FeynmanPF}) the expectation value is given by,
\begin{multline}
  \langle \sum_{i<j} \delta(r^{(i)}-r^{(j)}-r) \rangle = \frac{1}{\mathcal{Z}^{(N)}(\beta)} \frac{1}{N!} \sum_{[p]} \sigma_{[p]} M_{[p]} \int d{\bf R} \rho(r_{1},\dots,r_{N};\beta) \\
  \delta(r^{(i)}-r^{(j)}-r) \prod_{i=1}^{N} \frac{1}{\sqrt{4\pi\lambda'\beta}^{D}} \exp{[-\frac{(r_{[p](i)}-r_{i})^{2}}{4\lambda'\beta}]}
\end{multline}
We note that we only get a contribution to this sum if $i$ and $j$ are part of the same cycle. For each cycle length $l$, there are $C_{l} \binom{l}{2}$ equal contributions. Finally we note that the integral over the $\delta$-function gives an extra factor of $\sqrt{\frac{l}{l-1}}\exp{[-\frac{l r^{2}}{4(l-1)\lambda\beta}]}$. Thus we are left with,
\begin{eqnarray}
  \langle g(r) \rangle &=& \frac{2 \Omega}{N^{2}} \frac{1}{\mathcal{Z}^{(N)}(\beta)} \frac{1}{N!} \sum_{[p]} \sigma_{[p]} M_{[p]} \sum_{l=2}^{N} C_{[p]l} \binom{l}{2} \sqrt{\frac{l}{l-1}} \exp{[-\frac{l r^{2}}{4(l-1)\lambda'\beta}]} \prod_{l=1}^{N} P_{l}(\beta)^{C_{[p]l}} \\ 
                       &=& \frac{2 \Omega}{N^{2}} \sum_{[p]} \sigma_{[p]} \omega_{[p]} \sum_{l=2}^{N} C_{[p]l} \binom{l}{2} \sqrt{\frac{l}{l-1}}\exp{[-\frac{l r^{2}}{4(l-1)\lambda'\beta}]} \\
                       &=& \frac{2 \Omega}{N^{2}} \sum_{[p]} \sigma_{[p]} \omega_{[p]} \sum_{l=2}^{N} C_{[p]l} g_{l}(r;\beta)
\end{eqnarray}
Thus even though we have not done so in the main text, it is plausible that a similar reconstruction may be done for the pair correlation function as is done for the total energy and average value of the sign.


\input{fpimc-14-supplemental.bbl}
\bibliographystyle{aipauth4-1}
\end{document}